\documentstyle[sprocl]{article}

\def\Journal#1#2#3#4{{#1} {\bf #2}, #3 (#4)}

\def\NPB{{\em Nucl. Phys.} B}
\def\PLB{{\em Phys. Lett.}  B}
\def\PRL{\em Phys. Rev. Lett.}
\def\PRD{{\em Phys. Rev.} D}

\def\be{\begin{equation}}
\def\ee{\end{equation}}
\def\bea{\begin{eqnarray}}
\def\eea{\end{eqnarray}}

\begin{document}
\vbox{
\begin{flushright}
VAND-TH-98-12
\end{flushright}}

\title{The Warm Inflation Early Universe}

\author{Arjun Berera}

\address{Department of Physics and Astronomy \\ Vanderbilt University, 
Nashville, Tennessee 37235, USA\\E-mail: berera@vuhep.phy.vanderbilt.edu}

\maketitle\abstracts{  A general overview is given of the warm inflation
scenario.}

\bigskip

Talk presented at PASCOS-98, Northeastern University, March 1998

\bigskip


The thermodynamics of inflationary expansion can be either isentropic
\cite{guth,newi1,newi2} or
non-isentropic \cite{brsp,rudnei,wi,ab2}.  
Representing the early universe by a two fluid mixture of
radiation energy density $\rho_r$ and vacuum energy density
$\rho_v$, the inflationary regime, when the scale factor 
accelerates ${\ddot R} > 0$, is for
$\rho_v > \rho_r$.  From the point of view of
two fluid Friedmann cosmology, isentropic
inflationary expansion appears as a limiting case within the general regime
of non-isentropic of inflation.

In particle physics the vacuum equation of state $\rho_v=-p_v$ is realized by a
scalar field with energy density 
$\rho(\phi) = {\dot \phi}^2/2 + (\nabla \phi)^2 /2 +V(\phi)$, in which the
potential energy density dominates 
\begin{equation}
V(\phi) \gg \frac{1}{2} {\dot \phi}^2, \frac{1}{2} (\nabla \phi)^2.
\label{cond}
\end{equation}
Most field theory descriptions of inflation represent the vacuum energy through
a scalar field satisfying eq. (\ref{cond}), with $\phi$ referred to as the
inflaton.  The goal of inflationary scalar field dynamics is to sustain the
vacuum energy sufficiently long for expansion of the scale factor to exceed
observational lower bounds and then end the inflationary epoch by entering
the radiation dominated epoch.  In the context of dynamical models, isentropic
inflation is referred to as supercooled inflation and non-isentropic inflation
is referred to as warm inflation.

Scalar field models that represent the vacuum energy have no fundamental
motivation. They provide a convenient mode for studying the complex dynamics of
inflation. In addition, phase transitions can be represented by such models,
and that is another key element to the particle physics picture of the early
universe.

From the point of view of particle physics, a system of interacting fields will
exchange energy at all times, with the inflationary epoch having no a priori
reason for being different.  To insist otherwise imposes a sharp division
between a expansion regime with negligible energy exchange and then a subsequent
reheating period, which is essentially a second Big-Bang. 

On the one hand, for
a first-order phase transition with bubble nucleation kinetics, this is a
natural scenario dictated by the dynamics.  This conception of Guth's proved
unsuccessful \cite{guth,guthwein}
due to the conflicting requirements of having a slow nucleation rate to the true
vacuum for obtaining adequate expansion versus a fast rate to allow bubble
collisions after expansion, which would reheat the universe.

On the other hand, for continuous transitions, no qualitative feature of the
dynamics requires a sharp division between the expansion and heating periods.
Such a division can be constructed by requiring the scalar field potential to be
ultra-flat during the inflationary expansion period and then sharply cusped
to permit a subsequent reheating period that ends inflation and begins the
radiation dominated regime.  
This is the new inflation picture \cite{newi1,newi2}.
While the scalar field in on the ultra-flat
portion of the potential, it will have weak self-interaction, since such a
potential requires this. Also, it will interact weakly with other fields, 
since inflationary expansion will rapidly dilute any pre-existing field energy
apart from the vacuum energy.  An ultra-flat potential is sufficient for
satisfying the requirements of
inflation, but it is not necessary.

The general case is to allow the scalar field to interact with other fields
during the entire evolution down the potential well. With this follows reactive
forces on the inflaton, and they can slow the motion of the inflaton.  The
generic kinetics of continuous phase transitions for terrestrial systems,
Ginzburg-Landau kinetics, is emphatic on the dissipational properties of the
order parameter.   Cosmological theories of phase transitions assume that
the statistical mechanical principles
on such large scales are no different from those on
terrestrial scales.  Thus Ginzburg-Landau kinetics is a viable possibility for
the inflaton, provided that appropriate conditions can be realized.
In particular the dissipational dynamics of Ginzburg-Landau
kinetics assumes the presence of a large heat bath which interacts with the
order parameter.  For inflation, this requirement imposes 
the following self-consistency
condition. The inflaton must release adequate vacuum energy into the heat bath
to compensate for dilution of the heat bath due to inflationary expansion. Simultaneously
the reaction of the heat bath on the inflaton must be sufficient to slow the
inflaton's roll down the potential.  The questions are first can these
requirements be satisfied by a sensible phenomenological dynamics and second
can this dynamics be derived from quantum field theory?  

Consider stochastic evolution for the inflaton governed by the Langevin-like
equation 
\begin{equation}
{\ddot \phi}(t) + \left[ \Gamma + 3 \frac{\dot R(t)}{R(t)} \right]
{\dot \phi} + V'(\phi) = \eta(t)
\label{stoEOM}
\end{equation}
where $\eta(t)$ is a random force function with vanishing ensemble averaged
expectation value $\langle \eta(t) \rangle = 0$.  Consider the limit of strong
dissipation  $\Gamma \gg {\dot R}(t) /R(t)$ and
the overdamped regime 
\begin{equation}
\Gamma |{\dot \phi}| \gg |{\ddot \phi}|.
\label{odlim}
\end{equation}
Then eq. (\ref{stoEOM})  has the Ginzburg-Landau form
\begin{equation}
\frac{d \phi}{dt} = - \frac{1}{\Gamma} \frac{dV(\phi)}{d \phi}.
\label{overdamp}
\end{equation}
In this limit, the inflaton has a vacuum equation of state since
$\rho_{\phi} \approx V(\phi)$ which for a scalar field implies 
$p_{\phi} = -\rho_{\phi}$.  In addition, in a two fluid model
composed of the scalar field and radiation, by energy conservation
\begin{equation}
{\dot \rho}_r = -4 \frac{\dot R(t)}{R(t)} \rho_r - {\dot \rho}_{\phi}.
\label{rhoeom}
\end{equation}
The Friedmann cosmology determined by eqs. (\ref{overdamp}) and (\ref{rhoeom})
has been studied in \cite{ab2} for a variety of vacuum functions
$V(\phi) = \lambda M^{4-n} (M-\phi)^n$. It was found there that for
$2 \le n<4$ the scale factor will go from a radiation dominated behavior 
into an inflationary behavior and then smoothly back to a radiation dominated
behavior, and the latter occurring without reheating. 
The expansion e-folds, $N_e$, during inflation and the drop in $\rho_r(t)$ during the
inflationary period are determined by the index $n$ of the potential and the
dissipative coefficient $\Gamma$. For example for $n=2$, the quadratic limit,
$N_e =\sqrt{2\pi/(3\lambda)}(\Gamma/m_p)$ and 
$\rho_r(\tau_{EI})/\rho_r(\tau_{BI}) \approx 1/(4N_e^2)$, where the subscripts 
$BI$ and $EI$ signify begin and end inflation respectively.

The solutions in \cite{ab2} are an existence proof of the warm inflation 
regime.  A fundamental justification for such a dynamics requires deriving
equation (\ref{stoEOM}) from first principles and demonstrating the
consistency of the limit eq. (\ref{odlim}).
Eq. (\ref{stoEOM}) should not be confused with similar looking equations in
earlier reheating models \cite{reheat}, 
since in the warm inflation case, the dissipative term represents
frictional forces that arise from interaction of $\phi$ with the heat bath.
Furthermore, the overdamped limit eq. (\ref{odlim}) is equivalent to an adiabatic
limit. Under such conditions,
eqs. (\ref{stoEOM}) and (\ref{odlim}) are an outcome of quantum
mechanics both in flat spacetime \cite{cl} and for the cosmological setting of warm
inflation \cite{ab1}.  These equations have also been derived
from a particular quantum field theory
model in \cite{bgr}.

The fundamental origin of dissipation arises from the coupling of the
inflaton to other fields which comprise the heat bath.  In quantum field
theory this implies the interactions $\phi^2\chi^2$,
$\phi {\bar \psi} \psi$ and $\phi^2 A^{i\mu} A_{i \mu}$, 
for coupling to bosons $\chi$,
fermions $\psi$ and gauge fields $A_i^{\mu}$. For such couplings,
$\phi$ acts as a mass to the respective heat bath field and dissipation
effects are only relevant when the mass it induces is less than or of
order the temperature scale T.  To enhance dissipative effects for large
displacements of $\phi$, the couplings can also be modified by shifting
$\phi$ only in the relevant interaction term. For example, for the
bosonic case the shift 
$\phi^2 \chi^2 \rightarrow (\phi-M)^2 \chi^2$.
Thus for several heat bath fields a distributed mass model
suggests itself, which for the scalar case is
\begin{equation}
\sum_i g_i (\phi-M_i)^2 \chi_i^2
\end{equation}
or similarly a continuous mass model
\begin{equation}
\int d \mu g(\mu) (\phi-\mu)^2 \chi^2_{\mu},
\end{equation}
with the mass spectrum given by $g_i$, $g(\mu)$ respectively.
Such models could be motivated if the high energy world below the Planck
scale and well above the electroweak scale tended away from an organized
group theoretical structure towards a random one.

A second requirement of inflationary models is producing observationally
consistent density perturbations $\delta \rho/\rho \sim 10^{-5}$
\cite{newi1,newi2,bardeen,bf2,wi,mati}.  The general formula for 
$\delta \rho (t)/\rho (t)$ at horizon
entry $t_f$ for perturbations produced by a scalar field, that exited the
horizon at $t_i$ during inflation, is \cite{bardeen}
\begin{equation}
\frac{\delta \rho}{\rho} \equiv
\frac{\delta \rho}{\rho} (t_f) = 
\frac{V'(\phi(t_i)) \delta \phi(t_i)}{V(\phi(t_i))} \frac{1}{1+w}
\end{equation}
where $w \equiv p/\rho$. In the warm inflation case \cite{ab1}
\begin{equation}
\delta \phi^2(H_i{\bf {\hat e}},t_i) = \int_{V=1/H_i^3} \frac{d^3{\bf x}}{V}
e^{iH_i{\bf {\hat e}} \cdot {\bf x}} 
\langle \phi({\bf x}, t_i) \phi({\bf 0},t_i) \rangle_{\beta}
= O(1) \frac{H_i^3T}{V''(\phi(t_i))},
\end{equation}
where $H_i$ is the 
(slowly varying) Hubble parameter at time $t_i$ during warm inflation.
In warm inflation $\rho_r \gg {\dot \phi}^2/2$
so that
\begin{equation}
\frac{\delta \rho}{\rho} (t_f) = 
\frac{3V'(\phi(t_i)) \delta \phi(t_i)}{4\rho_r(t_i)}.
\end{equation}
In general the presence of non-negligible $\rho_r$ tends to suppress
$\delta \rho/\rho$. In models that have been examined, regions with
$\delta \rho / \rho <$ or $\ll 10^{-5}$ are much simpler to construct than
the opposite region of large amplitude.  In a phenomenological warm inflation
model \cite{wi} with a SU(5) 
Coleman-Weinberg potential, observational consistency has
been obtained  for both expansion e-folds $N_e > 60$ and density perturbations
$\delta \rho/\rho \sim 10^{-5}$.

In conclusion, the warm inflation scenario has 
been shown to be consistent with observation for
certain models and has nice features for treatment by quantum field theory.
The warm inflation regime also has interesting possibilities for production of
large scale cosmic magnetic fields
\cite{biermann} and baryogenesis \cite{ab2}, since in a sense, the warm
inflation regime is like a radiation dominated regime 
except with inflationary
expansion.


\section*{References}
\begin{thebibliography}{99}
\bibitem{guth} A. H. Guth, \Journal{\PRD}{23}{347}{1981}.
\bibitem{newi1} A. Albrecht and P. J. Steinhardt, 
\Journal{\PRL}{48}{1220}{1982}.
\bibitem{newi2} A. Linde, \Journal{\PLB}{108}{389}{1982}.
\bibitem{brsp} P. Spindel and R. Brout, 
\Journal{\PLB}{320}{241}{1994}.
\bibitem{rudnei} H. P. de Oliveria and R. O. Ramos, 
\Journal{\PRD}{57}{741}{1998};
A. V. Nesteruk, R. Maartens, and E. Gunzig, astro-ph/9703137.
\bibitem{wi} A. Berera, \Journal{\PRL}{75}{3218}{1995}.
\bibitem{ab2} A. Berera, \Journal{\PRD}{55}{3346}{1997}.
\bibitem{guthwein} A. Guth and E. Weinberg, 
\Journal{\NPB}{212}{321}{1983}.
\bibitem{reheat} A. Albrecht, P. J. Steinhardt, M. S. Turner
and F. Wilczek, \Journal{\PRL}{48}{1437}{1982};
A. D. Dolgov and A. D. Linde, \Journal{\PLB}{116}{329}{1982};
L. F. Abbott, E. Farhi, and
M. B. Wise, \Journal{\PLB} {117}{29}{1982}.
\bibitem{cl} A. O. Caldeira and A. J. Leggett, {\em Ann. Phys.} 
{\bf 149}, 374 (1983).
\bibitem{ab1} A. Berera, \Journal{\PRD}{54}{2519}{1996}.
\bibitem{bgr} A. Berera, M. Gleiser and R. O. Ramos, hep-ph/9803394.
\bibitem{bardeen} J. M. Bardeen, \Journal{\PRD}{22}{1882}{1980}.
\bibitem{bf2} A. Berera and L. Z. Fang, \Journal{\PRL}{74}{1912}{1995}.
\bibitem{mati} R. Maartens and D. Tilley, {\em Gen. Rel. Grav.}
{\bf 30} 289, 1998.
\bibitem{biermann} P. L. Biermann, H. Falcke, 
Proceeding of Frontiers in Contemporary
Physics International Lecture and Workshop, Vanderbilt University
1997; A. Berera,
talk at same conference. 




\end{thebibliography}
\end{document}